\documentstyle[aps,prd,preprint,tighten,floats,epsf]{revtex}

\def \GeV {{\rm GeV}}

\def \mb {{\rm mb}}
\def \mub {{\rm \mu b}}

\begin{document}
%

\preprint{PSU/TH/162}

\title{Double Pomeron Jet Cross Sections}

\author{Arjun Berera\thanks{E-mail: {\tt berera@phys.psu.edu}}
    and John C. Collins\thanks{E-mail: {\tt collins@phys.psu.edu}}
}

\address{
   Department of Physics,
   Pennsylvania State University,
   University Park, PA 16802, U.S.A.
}

\date {September 6, 1995}

\maketitle

\begin{abstract}
We treat hadron-hadron collisions where the final state is
kinematically of the kind associated with double-pomeron-exchange (DPE)
and has large transverse momentum jets.
We show that in addition to the
conventional factorized (FDPE) contribution, there is a
non-factorized (NDPE) contribution which has no pomeron beam jet.
Within a simple model we compute DPE-two-jet total and differential cross
sections at Tevatron energy scales, and show that the NDPE
contribution is dominant.
\end{abstract}


When two hadrons collide at high energy, there is about a
10\% chance that one of the incoming hadrons will survive
into the final state, while losing only a small fraction of its
energy.
Such events are called diffractive.
The object that is exchanged in diffractive processes is called the
``pomeron''. A primary characteristic is that cross sections with
pomeron exchange are approximately independent of energy at high
energy; the pomeron has spin close to unity \cite{pomeron}.
Moreover, there has been growing experimental evidence that within
diffractive events, hard QCD processes can also occur, i.e.,
processes with a large momentum transfer.  Such ``diffractive hard
scattering'' has been been seen not only in hadron-hadron scattering
by the UA1 and UA8 experiments \cite{UA1,UA8}, but also in
electron-proton scattering at the H1\cite{H1} and ZEUS\cite{ZEUS}
experiments, both in the deep-inelastic regime and in
photoproduction.

An interesting class of diffractive hadron-hadron collisions, is where
{\em both} hadrons survive unscathed, but leave a remnant system in
the central region of final-state rapidity.  Such events are called
double-pomeron exchange (DPE) events \cite {DPEthy,DPEexp}.  In
effect, one can try to consider such processes as pomeron-pomeron
collisions.  Our purpose in this paper is to explore the properties of
jet production in DPE events, and in particular the breakdown of
hard-scattering factorization.
Jet production by DPE has been reported \cite{DPE.jet.exp} in the UA1
detector.

The simplest model for diffractive hard scattering
is due to Ingelman and Schlein
\cite{IS}.  They treat each exchanged pomeron as a particle, in the
same manner as one obtains photoproduction cross sections from
electron-hadron scattering at small momentum transfer. Diffractive
hard scattering is then obtained by the use of parton densities in the
pomeron.  The model is used in much phenomenology \cite{H1,ZEUS}, and
has been applied to DPE\cite{DPE.IS}.  Interestingly, as Collins,
Frankfurt and Strikman (CFS) \cite{fs1,CFS} explained, the
Ingelman-Schlein model is not valid in QCD, even though the model
forms a useful phenomenological
benchmark \cite{CTEQ}.
CFS also predict a specific signature of the breakdown of
factorization, that at the leading twist level all the momentum of the
pomeron can go into the hard scattering---the process is
``super-hard'' or ``lossless''.
UA8 \cite{UA8} data gives experimental support to this result.

Our present paper shows how the mechanism of Collins, Frankfurt and
Strikman applied to DPE at the level of lowest-order Feynman graphs
provides a striking mechanism for the breakdown of factorization, and
we turn it into a quantitative, if crude,
model\footnote{%
   Preliminary results were presented in \cite{smallx}.
}
The model is in
effect a version of the Low-Nussinov-Gunion-Soper
model \cite{LowNussinov}, and the same model
was used by Berera and Soper
\cite{bersop} to understand properties of the pomeron's structure
function.  We will calculate the cross section for jet production in
DPE.  The important free parameter in our model is an overall
normalization which can be determined from elastic scattering.
Pumplin \cite{Pumplin} has done some calculations in the same
model, but without the strongly dominant gluon-gluon subprocess.
We call the model non-factorizing double-pomeron-exchange (NDPE), to
contrast it with the Ingelman-Schlein model applied to DPE, which we
call factorized double-pomeron-exchange (FDPE).

The dramatic feature of our model is that the pomerons have no beam
jets; the final state is then exceptionally clean, because it consists
of the two isolated, diffracted hadrons, the high transverse-momentum
jets, and nothing else.
Not only are such processes theoretically interesting in their own
right, but they have advantages for certain kinds of new particle
searches, provided the cross section is high enough, because a lot of
the background event is no longer present.  Production processes
of heavy flavors and Higgs are two such examples.
But studies of hard scattering in DPE should surely
start with the simplest processes, jet production.

Sch\"afer, Nachtmann and Sch\"opf \cite{SNS} have characterized the
final states we calculate as ``exclusive'' production , with the
Ingelman-Schlein process being a corresponding ``inclusive'' process.

Although there has been a significant amount of work on hard
processes with DPE, almost all of it has concerned production of heavy
quarks and Higgs bosons --- see, for example, Refs.~\cite{DPE.IS,SNS,BL}.
Moreover, much of it has used the Ingelman-Schlein method, the FDPE
process, which, as we will see, is much smaller than the NDPE process.
The
only work on the DPE-to-jets process that we know of is by Pumplin
\cite{Pumplin}.  He used the same model as we did, but restricted his
attention to the production
of quark jets.
As we will show in Sect.~\ref {sec.zero}, production of quark
jets is very much smaller than the production of
gluon jets, because it has a zero at zero momentum
transfer.  On the other hand, Pumplin paid more attention to good
modeling of the proton's elastic form factor.

Donnachie and Landshoff \cite{DL.model} have developed an extensive
phenomenology of a two-gluon model for the pomeron.
Bialas and Landshoff \cite{BL} used this model to calculate Higgs
production by DPE.  This model has many similarities to ours, and it
includes better modeling of the non-perturbative part of the
process.  In particular, it has a more precise treatment of the
pomeron-proton vertex and of the pomeron trajectory.

However, there is one important principle that our model establishes.
This is that the exclusive processes of DPE to jets is leading twist.
This is quite non-trivial, since in proving factorization, there are
some very non-trivial cancellations \cite{CSS}.
Indeed some of our graphs are a
power law larger than the final answer.  The proof of the necessary
cancellation relies on Ward identities applied to the hard part, as we
demonstrate around Eq.~(\ref{afterWI}) below.  To show that general
principles do not imply some other cancellation, it is important to
have a complete, consistent and gauge-invariant model.  This is
provided by our model, which consists of all the lowest-order Feynman
graphs that follow from a certain toy Lagrangian.

Our work also shows that one only has to be concerned about this kind
of cancellation in the hard scattering part of the graphs.  It also
indicates what principles need to be applied in a general process.
Without taking the cancellations into account, the calculation of the
Feynman graphs is much less efficient than it need be.  Of course,
once these results have been established, one is free to do more
precise modeling of the bound state.

Lu and Milana \cite {LuMilana} have also calculated Higgs production
by DPE, but they  did not realize that there is a
factorization-breaking mechanism.
Hence their cross section is too small, by orders of magnitude.

\section{Kinematics}
\label{sec.kin}

\begin{figure}
   \begin{center}
      \leavevmode
      \epsfxsize=0.25\hsize
      \epsfbox{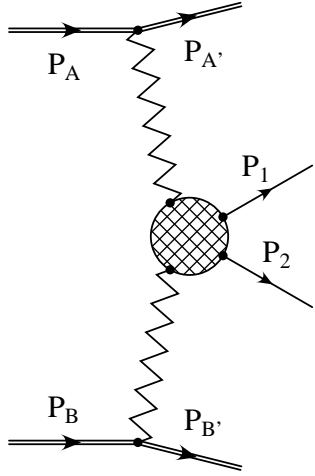}
   \end{center}
   \caption{
       Double pomeron exchange (DPE) to two jets.
   }
   \label{fig.DPE}
\end{figure}

The process we are interested in is
\begin{equation}A+B\rightarrow A'+B'+\mbox{\rm 2 jets}.
\label{DPE.jets}
\end{equation}
The hadrons $A$ and $B$ lose tiny fractions $x_a$ and $x_b$ of their
respective longitudinal momenta,
and they acquire transverse momenta ${\bf Q}_1$
and ${\bf Q}_2$.  (This defines a diffractive regime, and in Regge
theory would lead to an expectation of the dominance of
double pomeron exchange (DPE) --- Fig.~\ref{fig.DPE}.)
The jets carry large momenta of magnitude
$E_T$ in the plane perpendicular to the collision axis
with azimuthal angle $\phi$.  (This defines a hard-scattering
regime.)
The small transfer of
longitudinal momentum to the hard process implies large rapidity
gaps between the jets and the two outgoing hadrons.  In what follows,
we will generically denote the hadronic scale
by $m$.
We are interested in the kinematic
region where $m$, ${\bf Q}_1$, and ${\bf Q}_2$ are of a typical
hadronic scale (less than about
1 GeV), while $E_T$ is much greater than this scale, but
much less than $\sqrt s$.

To understand the asymptotics of Feynman graphs for
our process, it is convenient to use
light-cone components
$(+,-;{\bf \perp} )$.
The components of momenta of the hadrons
in Fig.~\ref {fig.DPE} are
\begin{eqnarray}
P_A&=&\left(\sqrt {\frac s2},\frac {M^2}{\sqrt {2s}};{\bf 0}\right
)\nonumber\\
P_B&=&\left(\frac {M^2}{\sqrt {2s}},\sqrt {\frac s2};{\bf 0}\right
)\nonumber\\
P_{A'}&=&\left((1-x_a)\sqrt {\frac s2},\frac {(M^2+{\bf Q}_1^2)}{
(1-x_a)\sqrt {2s}},{\bf Q}_1\right)\nonumber\\
P_{B'}&=&\left(\frac {(M^2+{\bf Q}_2^2)}{(1-x_b)\sqrt {2s}},(1-
x_b)\sqrt {\frac s2};{\bf Q}_2\right).\end{eqnarray}
We use bold-face type to indicate two-dimensional transverse
vectors.
For the jets, we ignore terms of relative order $\ll 1$ to
get:
\begin{eqnarray}
p_1&=&\left(ax_a\sqrt {\frac s2},bx_b\sqrt {\frac s2};E_T\cos\phi
,E_T\sin\phi\right)\nonumber\\
p_2&=&\left(bx_a\sqrt {\frac s2},ax_b\sqrt {\frac s2};-E_T\cos\phi
,-E_T\sin\phi\right),\end{eqnarray}
where it is convenient to define
\begin{eqnarray}
a&\equiv&\frac {1+\sqrt {1-\kappa}}2,\nonumber\\
b&\equiv&1-a,\end{eqnarray}
with
\begin{equation}\kappa\equiv\frac {4E^2_T}{x_ax_bs}.\end{equation}
We are interested in the limit where $x_a,x_b\rightarrow 0$, to give a
Regge-style limit, and where $E_T\to\infty$, to give a hard
scattering limit.  We hold $\kappa$, ${\bf Q}_1$, and ${\bf Q}_2$ fixed, to
correspond to a fixed angle for the hard scattering and fixed
transverse momenta for the outgoing hadrons.
In this limit, $s$ necessarily goes to infinity, since it is
proportional to $E_T^2/x_ax_b$.

\section{Model}

Our method is that of Berera and Soper \cite{bersop}.
We model the process by the lowest order Feynman
graphs that are appropriate.  The pomerons are replaced
by two-gluon exchanges, while instead of true bound
states, we model the hadrons by elementary color-singlet
scalar particles that we will call ``mesons'' and that are
coupled to scalar quarks by a $\phi^3$ coupling.

We normalize the coupling of our mesons so that we
reproduce the measured value for cross section for small
angle elastic scattering of protons, when we use
two-gluon exchange for elastic scattering.  This is
essentially the Low-Nussinov-Gunion-Soper model
\cite {LowNussinov}, but with some simplifications that
should not change the relative sizes of the cross
sections too much.  (We will address the adequacy of
our approximations in Sect.~\ref{conc}.)

\begin{figure}
   \begin{center}
      \leavevmode
      \epsfxsize=0.4\hsize
      \epsfbox{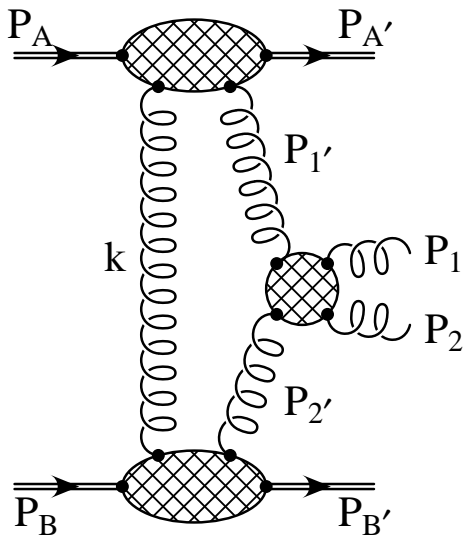}
   \end{center}
   \caption{
      Our model of the
      non-factorizing Double-Pomeron-Exchange (NDPE) amplitude
      with two gluon jets produced.
   }
   \label{fig.NDPEmodel}
\end{figure}

\begin{figure}
   \begin{center}
      \leavevmode
      \epsfxsize=0.4\hsize
      \epsfbox{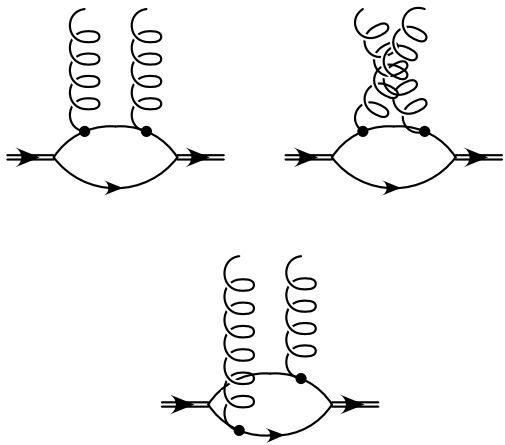}
   \end{center}
   \caption{
       Model for two-gluon form factor of meson.  There are three other graphs
identical to the above, except with the arrows on the quark lines reversed.
   }
   \label{fig.meson.glue}
\end{figure}

\begin{figure}
   \begin{center}
      \leavevmode
      \epsfxsize=0.4\hsize
      \epsfbox{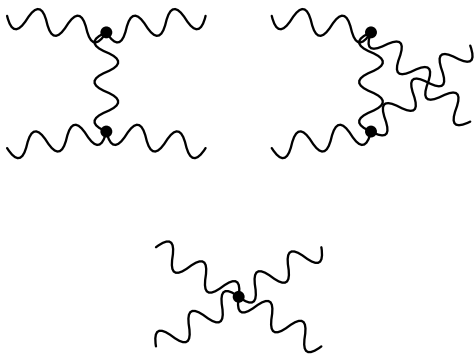}
   \end{center}
   \caption{
      Graphs for $g+g\to\mbox{\rm jet}+\mbox{\rm jet}$,
      when the jets are gluon
      jets, given that the jet pair is color-singlet.
      Similar graphs apply for production of a
      quark and antiquark jet.
   }
   \label{fig.gluon.jet}
\end{figure}

Our model for jet production by double pomeron exchange
(DPE) is therefore
given by Fig.~\ref {fig.NDPEmodel}.  Two gluons
couple to each of the diffracted hadrons:  this is the
minimum number of gluons that can couple to a
color-singlet.
The top and bottom bubbles imply a sum over all one
loop graphs with a quark loop --- Fig.~\ref {fig.meson.glue}.
One pair of gluons scatters to make the two jets, as in
Fig.~\ref {fig.gluon.jet}.  Note that since the mesons are
color-singlet,  the jet pair is color
singlet and hence Fig.~\ref {fig.gluon.jet} does not include the
graph with an $s$-channel gluon.
Also in Fig.~\ref {fig.meson.glue} the graph with a two-gluon
two-quark contact interaction is omitted, since, as we will see after
Eq.~(\ref{fppfmm}), this graph gives a non-leading contribution in
the Regge limit.  .

For our purposes, the most essential feature of the
pomeron is that it gives cross sections that are
approximately independent of the center of mass
energy $\sqrt {s}$, when
$\sqrt {s}$ is made large with all
transverse momenta fixed in size.  Such behavior is
given by exchanges of spin-1 fields --- see\cite {pomeron}.
Thus, in a perturbative model, the
lines coupling bubbles with very different rapidities
must all be gluons,
exactly as in Fig.~\ref {fig.NDPEmodel}.
Exchange of a quark-antiquark pair instead of a gluon
pair would give an amplitude suppressed by a power of
$s$.
Moreover, for
the same reason, the two-gluon form factors in
Fig.~\ref {fig.NDPEmodel}
must be between the meson
pairs $(A,A')$ and $(B,B')$, rather than between the pairs
$(A,B')$ and $(B,A')$.

\section{Calculation of Amplitude}

To compute the amplitude in Fig.~\ref {fig.NDPEmodel},
one first performs contour integrals for the longitudinal
components $k^{-}$ and $k^{+}$ of gluon loop momentum, while
making the appropriate approximations to get the leading
power behavior in the limit we are considering.  In the
leading region one finds \cite{IF}
\begin{eqnarray}
k^{\mu}&\sim&\left(m^2/\sqrt {s}, \, m^2/\sqrt {s}, \, m\right),\nonumber\\
p_{1'}^{\mu}&\sim&\left(x_a\sqrt s, \, m^2/\sqrt {s}, \, m\right),\\
p_{2'}^{\mu}&\sim&\left(m^2/\sqrt {s}, \, x_b\sqrt {s}, \, m\right).
\nonumber\end{eqnarray}
The following approximations can now be made:
\begin{itemize}
\item\ Neglect $k^{-}$ in all but the top bubble.
\item\ Neglect $k^{+}$ in all but the bottom bubble.
\item\ Take the dominant component ($+$ times $-$) in the
      product of the currents at the end of each gluon
      line.
\end{itemize}
In addition, the integral over all but one light-cone component of
momentum can be easily done for the loop integral inside
each form factor.  The integral
over the  transverse momentum
${\bf k}$ still links the different parts of the integral.
Since individual graphs for the hard scattering
Fig.~\ref {fig.gluon.jet} are a power $E_T^2/m^2$ bigger than
the sum, we cannot yet neglect the loop transverse
momentum in the hard scattering.

The approximations can be represented as follows.  Let
the exact formula for Fig.~\ref {fig.NDPEmodel} be
\begin{equation}\int \frac{d^4k}{(2 \pi)^4}\,
F^{\kappa\kappa'}(k,p_A,p_{A'})\,F^{\lambda
\lambda'}(-k,p_B,p_{B'})\,\,{\cal A}(\mu\nu ;f;p_{1'},p_{2'})\,\frac {
g_{\kappa\lambda}}{k^2}\,\frac {g_{\kappa'\mu}}{p_{1'}^2}\,\frac {
g_{\lambda'\nu}}{p_{2'}^2},\end{equation}
where $p_{1'} = k+p_A-p_{A'}$ and $p_{2'} = -k+p_B-p_{B'}$, while
$F$ represents the 2-gluon form factor of the
meson, Fig.~\ref {fig.meson.glue}, and ${\cal A}$ represents the
hard scattering subgraph, Fig.~\ref {fig.gluon.jet}.
The symbol $f$ labels the final state of the hard
scattering.
Then the approximations we have made are equivalent to
replacing this by
\begin{equation}
\int \frac{d^4k}{(2 \pi)^4}\,F^{++}((0,k^{-},{\bf 0}),p_A,p_{A'})
\,F^{--}((-k^{+}
,0,{\bf 0}),p_B,p_{B'})
{\cal A}(-+;f;p_{1'},p_{2'})\,\frac 1{-{\bf k}^2}\,\frac 1{-{\bf p}_{
1'}^2}\,\frac 1{{\bf p}_{2'}^2}.\label{fppfmm}\end{equation}
(The bold-face type indicates two-dimensional transverse
vectors.)
We now can see that, as stated earlier,
the graph where there is a two-gluon-two-quark
vertex in Fig.~\ref {fig.meson.glue}
gives a zero
contribution in  Eq.~(\ref{fppfmm}).
After doing the contour integrals, we find that the
momentum components of the gluons entering the hard
scattering are
\begin{eqnarray}
p_{1'}&=&\left(x_a\sqrt {\frac s2}+k^+,  \,
\frac {M^2}{\sqrt {2s}}-\frac {
M^2+Q_1^2}{(1-x_a)\sqrt {2s}} +k^-
; \, {\bf k}-{\bf Q}_{{\bf 1}}\right)
\nonumber\\
p_{2'}&=&
\left(\frac {M^2}{\sqrt {2s}}-\frac {M^2+Q_2^2}{(1-x_b)\sqrt {
2s}} -k^+
, \, x_b\sqrt {\frac s2} -k^-; \, -{\bf k}-{\bf Q}_{{\bf 2}}\right).
\end{eqnarray}
Within the hard scattering amplitude ${\cal A}$,
we will be able to replace
these vectors by $\left(x_a\sqrt s,0,{\bf 0}\right)$ and $\left(0
,x_b \sqrt s,{\bf 0}\right)$, but only after
taking care of the cancellations that reduce it by a factor of order
$m^2/E_T^2$.
The gluons $p_{1'}$ and $p_{2'}$ will be referred to as the incoming
gluons for the hard scattering.

The expressions we obtain for scattering amplitudes can conveniently
be written in terms of the following quantity obtained from
the meson form factor\footnote{%
   Compare the results for the impact factor in the
   theory of elastic scattering in \cite {IF}.
}
\begin{eqnarray}
\hat {g}_I({\bf k},{\bf Q})\equiv\frac {\delta^{ab}}{\sqrt {\strut
N_c^2-1}\sqrt {\strut({\bf k})^2}\sqrt {\strut({\bf Q}+{\bf k})^2}}\int_
0^1d\alpha\left\{G^{ab}_I(\alpha ,\alpha {\bf Q})-G^{ab}_I(\alpha
,{\bf k}+\alpha {\bf Q})\right\},
\label{ghat}
\end{eqnarray}
where $I$ labels the incoming hadrons ($A$ or $B$), $a$ and $b$ are
color indices,
$N_c=3$ is the number of colors,
and
where
\begin{equation}
G^{ab}_I(\alpha ,{\bf v})
=-g^2 G^2{\delta^{ab}}\frac {\alpha (1-\alpha
)}{(2\pi )^2|{\bf v}|}\frac 1{\sqrt {{\bf v}^2+4\Delta(\alpha)}}\ln\left
[\frac {\sqrt {{\bf v}^2+4\Delta(\alpha)}+{\bf v}^2}{\sqrt {{\bf v}^2+
4\Delta(\alpha)}-{\bf v}^2}\right].
\end{equation}
Here,
\begin{equation}
\Delta(\alpha)=m^2-\alpha (1-\alpha )M^2,
\end{equation}
and now $m$ represents the quark mass.  To obtain the integral in
Eq.~(\ref{ghat}) for $I=A$ we start with the sum of graphs in
Fig.~\ref {fig.meson.glue} (plus the three graphs with the quark lines
reversed). In accordance with the approximations that give
Eq.~(\ref{fppfmm}), we take the $+$ component for each of the vector
couplings, we set the $k^+=0$, and we apply
\begin{equation}
\int \frac{dk^-}{2\pi}.
\end{equation}
Finally we divide by $P_A^+$.

This gives the integral in Eq.~(\ref{ghat}) for the case $I=A$, and a
similar procedure is applied for ${\hat g}_B$.  The prefactors in
Eq.~(\ref{ghat}) are adjusted so as to absorb the color factors and
the denominators of the gluon propagators in a convenient way.

There is now one trick needed to simplify the
calculation of the amplitude Fig.~\ref {fig.NDPEmodel}, to exhibit the
cancellations we mentioned earlier.
We have used ${\cal A}(\mu ,\nu ;f)$ to represent the amplitude for
the hard scattering part, Fig.~\ref {fig.gluon.jet}.  Here, $\mu$
and $\nu$ are the Lorentz indices for the incoming gluons $p_{1'}$
and $p_{2'}$, and $f$ is a label for the jet part of the final
state.  As explained above, we need only the
component ${\cal A}(-,+;f)$.  However, individual
Feynman graphs for ${\cal A}$ are a factor of order $E_T^2/m^2$ bigger than
the final answer.  Following standard methods, we add
terms that are zero by the QCD Ward identities and obtain:
\begin{equation}{\cal A}(-,+;f)={\cal A}(-,+;f)-\frac {p_{1'\mu}{\cal A}
(\mu ,+;f)}{p_{1'}^{+}}-\frac {{\cal A}(-,\nu ;f)p_{2'\nu}}{p_{2'}^{
-}}+\frac {p_{1'\mu}{\cal A}(\mu ,\nu ;f)p_{2'\nu}}{p_{1'}^{+}p_{
2'}^{-}}.\end{equation}
After dropping terms that are non-leading by a power,
we obtain
\begin{equation}{\cal A}(-,+;f)=\frac {p_{1'i}{\cal A}(i,j;f)p_{2'
j}}{p_{1'}^{+}p_{2'}^{-}},\label{afterWI}\end{equation}
where the latin indices $i$ and $j$ refer to the transverse
components of vectors.
In Eq.~(\ref{afterWI}), the factor $p_{1'i}p_{2'j} /
(p_{1'}^{+}p_{2'}^{-})$ is of order $m^2/E_T^2$, which demonstrates the
previously mentioned cancellation.
Within the hard scattering amplitude ${\cal A}(i,j;f)$,
it is now correct to approximate
$p_{1'}$ by $(x_a\sqrt{s/2},0,{\bf 0})$
and $p_{2'}$ by $(0,x_b\sqrt{s/2},{\bf 0})$.

For the amplitude we now find that
\begin{eqnarray}
{\cal M}&=&-\frac {(-i)}{x_ax_b}\int\frac {d^2{\bf k}}{(2\pi )^2}\,
\hat {g}_A({\bf k},-{\bf Q}_1)\,\hat {g}_B({\bf k},{\bf Q}_2)\,\epsilon_
i({\bf k}{\bf -}{\bf Q}_{{\bf 1}})\,\epsilon_j(-{\bf k}{\bf -}{\bf Q}_{
{\bf 2}})\,{\cal A}(i,j;f)\nonumber\\
&\equiv&\frac {-i B_{ij}({\bf Q}_1,{\bf Q}_2)\,{\cal A}(i,j;f)}{x_ax_
b} .\nonumber\\
\label{mdp}\end{eqnarray}
Here, the ``polarization" vectors are defined as
\begin{equation}\epsilon_i({\bf k})=\frac {{\bf k}_i}{\sqrt {{\bf k}^
2}}.\label{pol}\end{equation}
It will be convenient to define
\begin{equation}\overline {{\cal M}}=x_ax_b{\cal M}.\end{equation}
This scaled amplitude has the property of being
independent of $x_a$, $x_b$, and $E_T$ in the combined Regge and
hard scattering limit that we are considering.

We note for later use that when the
transverse momenta of the two outgoing hadrons are
zero, the hadronic tensor $B_{ij}$ in Eq.~(\ref {mdp}) is invariant
under rotations about the collision axis, so that we can
write
\begin{equation}B_{ij}({\bf 0},{\bf 0})= 2\delta_{ij}{\beta}^
2(0).\label{bat0}\end{equation}
(We will see later that ${\beta}$ is related to the
meson-pomeron coupling in our model; its definition here
agrees with Eq.~(\ref {bet2}) below.)
It will also be convenient to extract the couplings and
to define
\begin{equation}\tilde {B}_{ij}=\frac {B_{ij}}{g^4G^4}.
\label{Btilde.def}
\end{equation}

In Eq.~(\ref {mdp})
${\cal A}(i,j;f)$ is the amplitude
for an
incoming gluon pair in linear polarization state $i,j$ to
go to the outgoing final state $f$.
The incoming gluon pair for the
hard scattering will be in an overall color
singlet state since the remaining portion of the final
state, the two mesons, is color singlet.
We will refer to ${\cal A}(i,j;f)$ and
$B_{ij}({\bf Q}_1,{\bf Q}_{{\bf 2}})$ as the hard scattering amplitude
and the hadronic amplitude, respectively.
Both amplitudes are $2 \times 2$
tensors in the transverse space of gluon polarizations.

\section{Cross Section}

We now give formulae for the cross section.  In terms of the
light-cone momentum fractions $x_a$ and $x_b$,
\begin{equation}x_ax_b\frac {d\sigma}{d^2{\bf Q}_1d^2{\bf Q}_2dE^
2_Td\phi_jdx_adx_b}=\frac {|\overline {{\cal M}}|^2\kappa^2}{2^{1
7}\pi^8E_T^4\sqrt {1-\kappa}} \label{dsig1} \end{equation}
with implicit sums over final state color, flavor and spin.
Next let us
change to the rapidity variables of the jet partons
in the final state.
These are defined as
\begin{equation}
y_1\equiv\frac 12\ln\frac {p_1^{+}}{p_1^{-}},~y_2\equiv\frac
12\ln\frac {p_2^{+}}{p_2^{-}},\end{equation}
so that
\begin{equation}\kappa =\frac 1{\cosh^2(\frac {y_{-}}2)},\end{equation}
where $y_- \equiv y_1-y_2$.
Note that $|\overline {{\cal M}}|^2$ depends only on $y_{-}$.  In terms of
this variable and $y_{+}\equiv (y_1+y_2)/2$ we have
\begin{equation}\frac {d\sigma}{d^2{\bf Q}_1d^2{\bf Q}_2dE^2_Td\phi_
jdy_{-}dy_{+}}=\frac {|\overline {{\cal M}}|^2\kappa^2}{2^{17}\pi^
8E_T^4}. \label{dsig2} \end{equation}

The squared amplitude can be expressed in terms of
the hadronic and hard amplitudes as follows:
\begin{equation}|\overline {{\cal M}}|^2=B^{*}_{ij}H_{ijkl}B_{kl},
\label{md2}\end{equation}
where $B_{ij}$ is defined in Eq.~(\ref {mdp}) and
\begin{equation}H_{ijkl}(\kappa )=\sum_f{\cal A}^{*}(i,j;f)
{\cal A}(k,l;f).\label{H}\end{equation}
In the hard amplitude ${\cal A}(i,j;f)$, $i$ and $j$ are gluon polarization
indices while $f$ generically represents any final parton pair state.
The sum $\sum_f$ is over all spin, flavor and color states for
final-state partons of given momenta.

{}From Eqs.~(\ref{dsig1})--(\ref {md2})
at fixed $\kappa$, we see that the differential cross
section is leading twist and has the
appropriate large $s$ behavior to approximate an amplitude
with pomeron exchange.  That is, when $E_T$ gets large
and $x_a$ and $x_b$ get small, the
amplitude $\overline {{\cal M}}$ is constant,
and the cross section goes like $1/E_T^4$.

The easiest way to see that this is a leading twist
contribution is to verify that the same power laws are
obtained from photon exchange.  Such a process is
obtained when one replaces the hadrons in reaction
(\ref {DPE.jets}) by charged particles, and the exchanged
pomerons in Fig.~\ref {fig.DPE} by single photons.
Normally, when one has a short-distance subprocess
initiated by elementary particles like photons, and one
replaces the photons by a composite objects like
pomerons, one loses a power of the scale $E_T$ of the hard
scattering.  But this result fails when a diffractive
requirement is imposed on the final state, as explained by
Collins, Frankfurt and Strikman \cite{CFS}.  Our model is
perhaps the simplest example of a failure of
hard-scattering factorization.
We have exhibited all the
graphs at a certain order of perturbation theory for a particular
final-state, and
there is no possibility of cancellation.  The
factorization theorem is an order-by-order result; to get
factorization one has a cancellation between the graphs
that
we calculate and some other graphs with different final states.

\section{Normalization from Elastic Scattering}
\label{secelas}

The parameters of our model are:  the mass $m$ of the
scalar quark, the mass $M$ of the mesons, the
quark-meson coupling $G$, and the gauge coupling $g$.  We
suppose that the most important unknown is the
normalization of our model for the non-perturbative
physics.  So we set all the masses to a typical hadronic
scale:  $m^2=M^2=0.1~\GeV^2$.  As for the couplings,
we have a factor $g^2$ in the hard scattering amplitude ${\cal A}$
and a factor $g^4G^4$ in the hadronic part.  The $g^2$ in
the hard amplitude should be given by the usual running
coupling at
a scale of order $E_T$, and we now
determine a suitable numerical value for $g^4G^4$ from
applying our model to elastic scattering.  This model is
effectively the Low-Nussinov model \cite{LowNussinov}.

We write the elastic-scattering amplitude for our model
in the Regge-like form
\begin{equation}{\cal M}_{\rm el}(t)=s\beta^2(t).\end{equation}
Here, the pomeron-hadron coupling is
given by a transverse integral,
\begin{eqnarray}
\beta^2(t)&=&-\frac{1}{4}g^4\int\frac {d^2{\bf k}}{(2\pi )^2}\hat {g}_A({\bf k}
,{\bf Q})\hat {g}_B({\bf k},{\bf Q})%
\label{bet2},\end{eqnarray}
where ${\bf Q}$ is the momentum transfer, so that
$t\equiv -{\bf Q}^2$, and $\hat {g}({\bf k},{\bf Q})$ is defined in
Eq.~(\ref {ghat}).
We also define %
\begin{equation}\tilde{\beta }(t)^2\equiv\frac {\beta (t)^2}{g^4G^
4}.\label{betti}\end{equation}

The elastic scattering cross section is then
\begin{equation}\left(\frac {d\sigma (t)}{dt}\right)_{\rm el}=\frac {
|{\cal M}_{\rm el}|^2}{16\pi s^2}=g^8G^8\frac {|\tilde{\beta }(t)|^4}{
16\pi}.\label{dsig}\end{equation}
We can now determine $gG$ by equating
the left-hand side of this equation to
the pp-elastic cross section at some particular $t=t_0$ and
at some appropriate value of $s$.
{}From Eq.~(\ref {dsig}) we can reexpress the DPE
amplitude $|\overline {{\cal M}}|^2$ (defined in
Eqs.~(\ref {mdp}) -- (\ref {Btilde.def}))
as follows:
\begin{equation}|\overline {{\cal M}}|^2=16\pi\left(\frac {d\sigma (t_
0)}{dt}\right)_{\rm el}\frac {\tilde {B}^{*}_{ij}H_{ijkl}\tilde {B}_{
kl}}{|\tilde{\beta }(t_0)|^4},\label{m2}\end{equation}
where the unknown couplings have dropped out.

This equation, (\ref {m2}), is particularly simple for
forward scattering of $A$ and $B$, when we set all the
momentum transfers to zero:
$Q_1=Q_2=t_0=0$.  Then
we use Eq.~(\ref{bat0}) to give
\begin{equation}|\overline {{\cal M}}(0,0)|^2=64\pi\left(\frac {d\sigma
(0)}{dt}\right)_{\rm el}
\delta_{ij}\delta_{kl}H_{ijkl},\label{forcx}\end{equation}
which relates the DPE-to-jets amplitude to an elastic
cross section and a hard scattering amplitude
The elastic cross section is observable, and hard
scattering amplitudes are perturbatively calculable.

\section{Numerical Results}
\label{seccrx}

In this section we compute numerical values for the
DPE-to-jets cross section.

\subsection{Zero in forward quark-jet production}
\label{sec.zero}

Now, there are two hard
subprocesses:  $gg\to gg$ and $gg\to q\bar {q}$.  We first show that the
$gg\to q\bar {q}$ subprocess gives a zero in its contribution to the
DPE-to-jets process when the transverse momentum
transfers from the hadrons are zero, a result previously
obtained by Pumplin \cite {Pumplin}.  Since the $gg\to gg$
gives no such zero, it must be by far the dominant
subprocess, and Pumplin's estimates for the cross
section should be much lower than the true values.

By a straightforward
perturbative calculation, one can verify that for the quark amplitude,
$gg \rightarrow q\bar{q}$,
\begin{equation}{\cal A}_q(1,1,\lambda_1,\lambda_2)=-{\cal A}_q(2
,2;\lambda_1,\lambda_2)\end{equation}
where $\lambda_i$ is the helicity of quark $i$.  This property
implies, from Eq.~(\ref {forcx}),
that the quark amplitude vanishes when
the hadrons $A$ and $B$ are forward scattered.  That this
same cancellation does not occur for the gluon amplitude
can be verified from Eq.~(\ref {ampls}) in the appendix.

The cancellation in the case of quark jet production can
be understood by examining the helicity amplitudes given
by Gastmans and Wu \cite {GW} for $e^{+}e^{-}\to\gamma\gamma$.  (Notice
that because our final state is color-singlet, only the
abelian parts of the graphs are relevant.)

Hence our cross section is dominated by the production
of gluon jets.  All the polarization combinations for
the $gg\rightarrow gg$ amplitude are given in the appendix, in
Eq.~(\ref {ampls}).
Special attention is given in deriving these amplitudes
with incoming linearly polarized gluons.
In terms of the amplitudes
in the appendix, the final state sum in Eq.~(\ref {H})
requires a polarization sum and a factor of 8 for color:
\begin{equation}H_{ijkl}=8\sum_{\alpha \beta }{\bar A}^{*}
(i,j;\alpha,\beta){\bar A}(k,l;\alpha,\beta).
\label{hndpe}\end{equation}

\subsection{Comparison cross sections}

\begin{figure}
   \begin{center}
      \leavevmode
      \epsfxsize=0.3\hsize
      \epsfbox{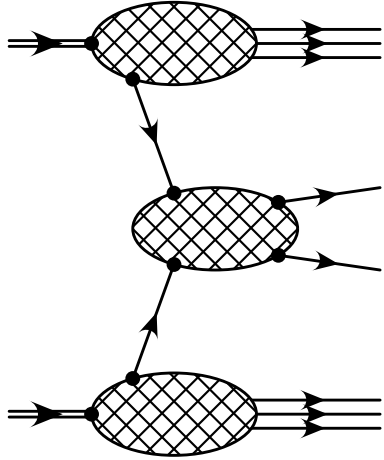}
   \end{center}
   \caption{
      Amplitude for inclusive two-jet production.
   }
   \label{fig.2jet.incl}
\end{figure}

\begin{figure}
   \begin{center}
      \leavevmode
      \epsfxsize=0.3\hsize
      \epsfbox{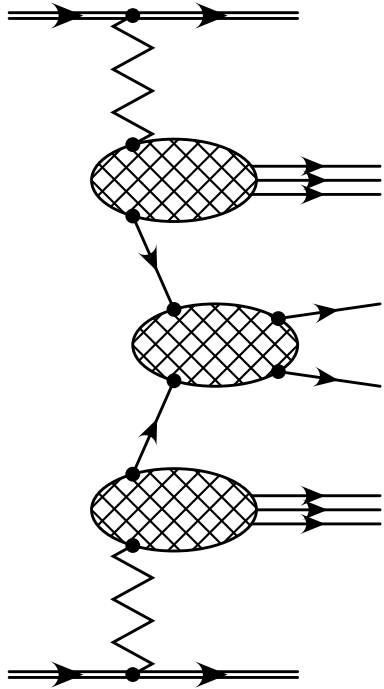}
   \end{center}
   \caption{
       Factorized-Double-Pomeron-Exchange (FDPE) amplitude
       with two jets produced.
   }
   \label{fig.FDPE}
\end{figure}

For comparison purposes, we have also computed
(a) the
inclusive two-jet cross section (Fig.~\ref {fig.2jet.incl})
(i.e., without a diffractive requirement:
$A+B\to\mbox{\rm jet}+\mbox{\rm jet}+X)$, and
(b) the result of applying the
Ingelman-Schlein model to DPE, which gives a result for the
process $A+B\to\mbox{\rm $A'+B'+$jet}+\mbox{\rm jet}+X.$ This last process we
call the factorized double-pomeron-exchange (FDPE)
two-jet cross section, and it is given by graphs like
Fig.~\ref {fig.FDPE}.

For inclusive two-jet production, the differential cross section is,
\begin{equation}\frac {d\sigma}{dE^2_Tdy_{-}dy_{+}}=\sum_{a,b}x_a
f_{a/A}(x_a)x_bf_{b/B}(x_b)\frac {\kappa^2H^{ab}_{\rm incl}(\kappa )}{
256\pi E^4_T},\end{equation}
where
\begin{eqnarray}
x_a&=&\frac {E_T}{\sqrt {s}}(e^{y_1}+e^{y_2}),\nonumber\\
x_b&=&\frac {E_T}{\sqrt {s}}(e^{-y_1}+e^{-y_2}),\end{eqnarray}
and $H^{ab}_{\rm incl}$ is the standard hard inclusive matrix
element squared for incoming partons $a$, $b$.  It is summed
over all final parton pairs.  Note that it is
averaged over initial polarization
states of the incoming partons $a$ and $b$.

For FDPE, the natural extension of the
Ingelman-Schlein model
\cite {IS} gives
\begin{eqnarray}
\frac {d\sigma}{dE^2_Tdy_{-}dy_{+}}&=&\frac {\kappa^2}{256\pi E^4_
T}\left(\frac N{16\pi}\right)^2\int dx_{P_A}dx_{P_B}B^A_{t_A}(x_{
P_A})B^B_{t_B}(x_{P_B})\nonumber\\
&&x_af_{a/P_A}(x_a/x_{P_A})x_bf_{b/P_B}(x_b/x_{P_B})H^{ab}_{\rm incl}
(\kappa ),\end{eqnarray}
where $N$ is a normalization factor \cite{CTEQ} that we
have set equal to $2/\pi$ to give the conventions of
Donnachie and Landshoff \cite{DL.norm}.  We have used
the following integral over the momentum transfer:
\begin{eqnarray}
B^I_t(x_P)&=&\int^0_{-\infty}dt\,\,|\beta_I(t)|^2x_P^{-2\alpha_{{\bf P}}
(t)}\nonumber\\
&=&\sigma_0x_P^{-2\alpha_0}\int^0_{-\infty}dt\,e^{(b_0+2\alpha'\ln
x_P)t},\end{eqnarray}
where $\alpha_{{\bf P}}(t)\equiv\alpha_0+\alpha't$ is
the pomeron trajectory function.
We will use
$\sqrt {\mathstrut\sigma_0}=4.6~\mb^{\frac 12}$,
$b_0=3.8~\GeV^{-2}$,
$\alpha_0=1.08$ and $\alpha'=0.25~\GeV^{-2}$, following \cite {DL}.

\subsection{Results}

\begin{figure}[p]
   \begin{center}
      \leavevmode
      \epsfxsize=0.8\hsize
      \epsfbox{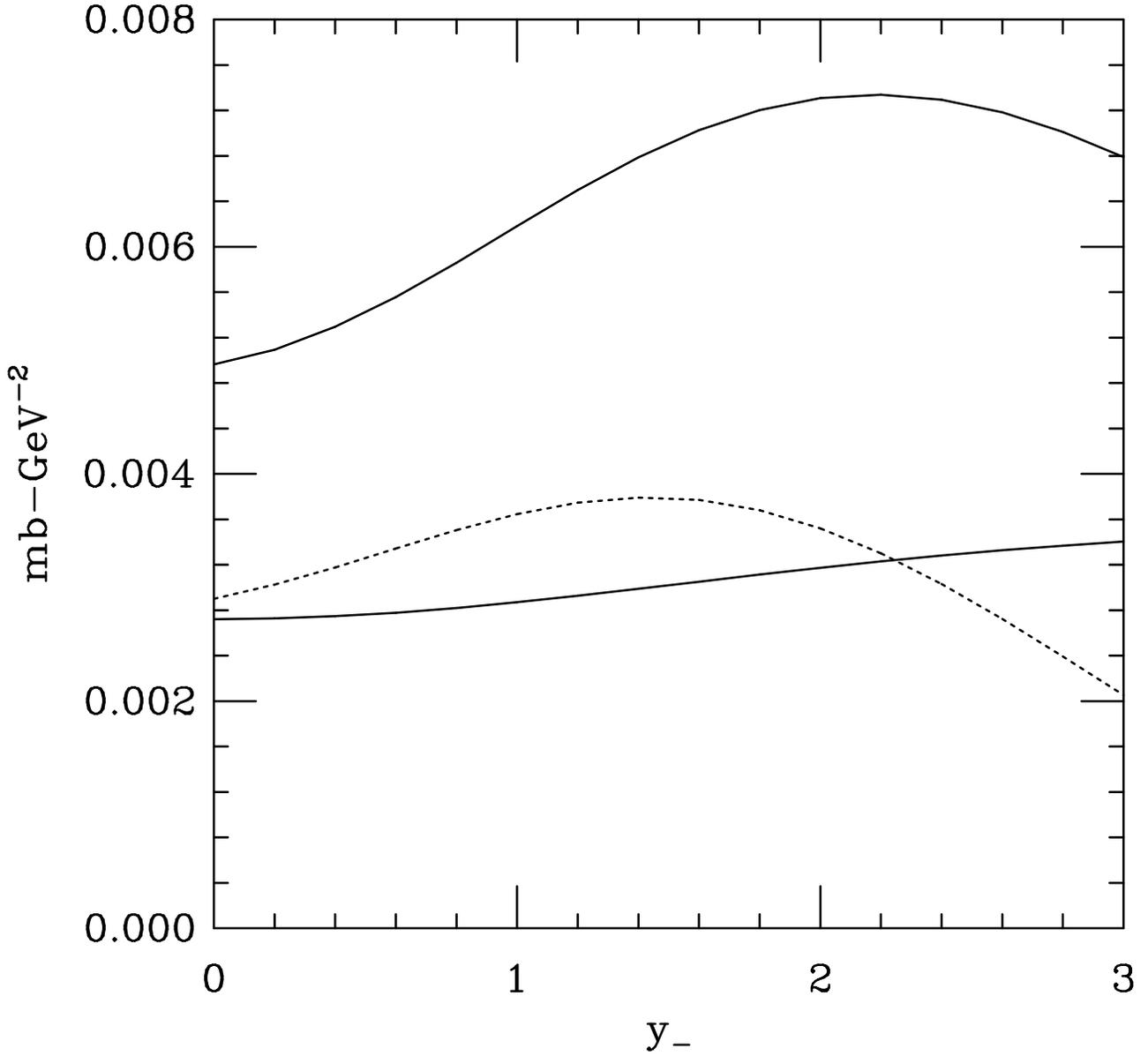}
   \end{center}
   \caption{
      Differential cross section $d\sigma/dE^2_T dy_+dy_-$
      at $\protect\sqrt{s}$=1800 GeV, $E_T$=5 GeV
      for: top solid two-jet inclusive, bottom solid 2 X (NDPE),
      and dashed 200 X (FDPE).
   }
   \label{fig.diff.xsect}
\end{figure}

In Fig.~\ref {fig.diff.xsect}, the differential cross section,
$d\sigma /dE^2_Tdy_{+}dy_{-}$, is plotted in $\mb/\GeV^2$
at $\sqrt {s}=1800~\GeV$, $y_{+}=0$ and $E_T=5~\GeV$.  We give the
differential cross section for the three cases: inclusive jet
production, for factorized DPE, and for
our model of non-factorizing DPE.

In the calculation of inclusive cross sections, the parton
distributions in the proton
are those of CTEQ1 \cite{cteq1}.
In the calculation of the FDPE cross section, the parton
distributions in the Pomeron are given by using the
CTEQ package to evolve the distributions numerically from
initial distributions at $Q^2=4~\GeV^2$.
As initial distributions, we use an ansatz like that
of Ingelman and Schlein \cite {IS}.  For the gluons we
choose
\begin{equation}f_{g/P}(x)=3(1-x),\end{equation}
and for the light quarks we choose
\begin{equation}f_{u/P}(x)=f_{d/P}(x)=\frac 34(1-x);\end{equation}
we set all other parton species to zero.  These initial
distributions satisfy the momentum sum rule, and split
the pomeron's momentum equally between gluons and
quarks.  For our estimates of cross sections, these
parton densities are sufficiently close to fits obtained at
HERA \cite{ZEUS}.
In all our calculations,
the distribution functions were evolved with three quark flavors
and $\Lambda =0.281 ~\GeV$.  In the hard amplitude, the
running coupling constant was evaluated at $E_T$.

For NDPE, we normalize the couplings in our model of
the hadrons from the forward elastic scattering cross
section, as explained in Sect.~\ref {secelas}.
We use the value $(d\sigma (0)/dt)_{\rm el}=201~\mb/\GeV^2$, which is
obtained by the optical theorem from a total cross
section $\sigma^{\bar {p}p}_{\rm tot}=62.7~\mb$ \cite {DL,ua4,bozzo}.
Since the true elastic cross section
depends on energy, but our model's cross section does
not, we chose to use the measured cross section
at $\sqrt {s}=540~\GeV$.
so that the
pomeron is being sampled under approximately the same
conditions in both the DPE-to-jets and the elastic
scattering process.
The NDPE cross section depends
linearly on the elastic cross section---see
Eq.~(\ref {m2})---
so that one can trivially
rescale our predicted cross sections when a different
value for $(d\sigma (0)/dt)_{\rm el}$ is preferred.

The total cross
sections integrated over $y_{+}$, $y_{-}$ and for $E_T>5.0~\GeV$
are, at $\sqrt {s}=1800~\GeV$,
\begin{eqnarray}
\sigma_{\rm incl}(1800,5)&=&2.4~\mb,\nonumber\\
\sigma_{\rm FDPE}(1800,5)&=&0.0022~\mb,\nonumber\\
\sigma_{\rm NDPE}(1800,5)&=&0.68~\mb,\nonumber\\
\label{sigma.1800}\end{eqnarray}
and at $\sqrt {s}=630~\GeV$,
\begin{eqnarray}
\sigma_{\rm incl}(630,5)&=&0.31~\mb,\nonumber\\
\sigma_{\rm FDPE}(630,5)&=&0.000062~\mb,\nonumber\\
\sigma_{\rm NDPE}(630,5)&=&0.18~\mb.\nonumber\\
\label{sigma.630}\end{eqnarray}
The integration range for the rapidities $y_{+}$ and $y_{-}$ was restricted
so that both incoming partons into the hard process had momentum fractions
less than 0.05.
This corresponds to
a typical selection cut on the diffractive
hadron for identifying pomeron exchange events.
The same
cut on the incoming partons was made for the estimate of
the inclusive jet cross sections.  If one allows the complete range of
momentum fractions from 0 to 1 for this case one obtains
$\sigma_{{\rm i}{\rm n}{\rm c}{\rm l}}(1800,5)=5.0~\mb$
and $\sigma_{{\rm i}{\rm n}{\rm c}{\rm l}}(630,5)=1.3~\mb$.
For quark jets in the NDPE case, we have verified that their cross sections
are at least a few orders of magnitude smaller.
Also for the NDPE case,
by a simple hand calculation of the total
cross section using Eqs.\ (\ref{dsig2}) and (\ref{forcx})
one can confirm the
order of magnitudes quoted above.  From this
exercise, the factor 4 increase at
$\sqrt{s}=1800 ~\GeV$ compared to $\sqrt{s}= 630 ~\GeV$
is seen to arise primarily from the same factor increase in the
accessible region of jet rapidity.

We will comment on the implications of these calculations in the next
section.

\section{Conclusion}
\label{conc}

We have estimated jet production in DPE processes, by a mechanism in
which the whole of the momentum of the pomerons goes into the jets.
The process has a quite dramatic signature: the final state consists
of the two diffracted hadrons, two high-$E_T$ jets, and {\em
nothing} else.  It is a leading-twist contribution, not suppressed
by a power of $E_T$, despite the fact that the pomeron is a composite
object.  This is permitted because the factorization theorem does not
apply to diffractive processes.

Moreover, the cross section is an order of magnitude larger or more
than the obvious Ingelman-Schlein type process.  The lack of pomeron
beam jets is presumably the reason why we are able to get such large
cross sections.

Our model is, of course, quite primitive.  However, by
being a complete and consistent calculation in lowest
order perturbation theory, it does establish an important
principle:  that one cannot treat the exchanged pomeron
as a particle with associated parton densities.  The
approximations we have made are concerned only with
extracting the leading power behavior in the appropriate
kinematic limit.
Since gauge-invariance causes
some quite non-trivial cancellations before the final
result is obtained, the exact gauge invariance and
consistency of our model is critical to establishing the
general principles.

We have also a new relation, Eq.~(\ref{forcx}), which
relates the DPE-to-jets cross section at zero momentum
transfer to the elastic scattering cross section.  This
relation is independent of all details of the hadronic
form factor, and depends only on the use of a
two-gluon-exchange model.

There are at least two obvious sources of large corrections to our
model:
\begin{itemize}
\item\ Absorptive corrections due to extra exchanges of
   pomerons and gluons between those objects in our model
   that have very different rapidities.  Recent work by
   Gotsman, Levin and Maor \cite {GLM} gives one way
   of tackling this problem.
\item\ Sudakov corrections that suppress the process of
   two gluons of low virtuality going into a hard process.
   Standard QCD methods can presumably be applied here.
\end{itemize}
{}From Ref.~\cite{GLM}, for example, we
expect that the true DPE-to-jets cross section will be an
order of magnitude or more smaller than the values we
have calculated.
But remember that absorptive corrections will reduce all DPE cross
sections, including FDPE.
More precise modeling of the hadron form
factor is also needed.  But this is presumably a less
important matter.

We plan to return to these issues in a later paper.

In addition, a comparison with existing data \cite{DPE.jet.exp} from
the UA1 detector would be useful.  Since more experimental and
phenomenological work is necessary to obtain correctly normalized
cross sections, we will merely note here some orders of
magnitude.  At $\sqrt s=630~\GeV$, Ref.~\cite{DPE.jet.exp} quotes a
cross section for detected DPE events of $0.3~\mub$, whereas Streng's
\cite {DPE.IS} estimate for the same quantity is quoted as $5~\mub$.
This discrepancy is attributed to the stringent cuts imposed on the
data.

Moreover, the experiment finds that $49\%$ of their events have one
or more jets with $E_T$ above $5~\GeV$.
Normally, in a hadron-hadron collision,
such a jet fraction would be impossibly high.  For example,
our FDPE cross section, in Eq.~(\ref {sigma.630}), after reduction by
absorption and experimental cuts, is much smaller than the data.
But our
NDPE cross section is tens of $\mub$, so until we apply absorption and
Sudakov corrections this partial cross section is in danger of being
larger than the full DPE cross section.

What we may reasonably conjecture is that hard scattering cross
sections in DPE may be surprisingly high.  Although our cross sections
are about
an order of
down from the fully inclusive jet
cross section (with the same cuts), it should be remembered that the
DPE cross section (without the jet requirement) is itself much smaller
than the total cross section \cite {DPEthy,DPE.IS}.

We suggest that a qualitative understanding of the large
cross sections can be obtained by observing that, in our
model, Fig.~\ref {fig.NDPEmodel}, we really do not have
pomerons exchanged between the hadrons and the jets,
but only single gluons.  So that we get a color singlets
coupling to the hadrons, it is sufficient to add one extra
gluon exchanged between the two diffracted hadrons.

If further examination supports the sizes of cross sections we
predict, then the mechanism we are using could be very important for
all kinds of studies.  For example, one can produce the Higgs boson
\cite{BL}.
Although the cross section would be a lot lower than the total Higgs
cross section, the lack of a background event could make up for the
lower rate in terms of the usefulness of the signature, at least for
certain ranges of parameters.  Compare Ref.~\cite {Bj}.

\section*{Acknowledgments}

We thank M. Albrow, L. Frankfurt, K. Goulianos,
D. Soper, and M. Strikman for helpful comments.
This work was supported in part by the U.S. Department
of energy under grant numbers DE-FG02-90ER-40577 and
DE-FG02-93ER40771.

\appendix
\section*{Formulas for Amplitudes}

We give the amplitudes for
\begin{equation}g(p_{1'},\epsilon_{1'}(i))+g(p_{2'},\epsilon_{
2'}(j ))\rightarrow g(p_1,\epsilon_1({\alpha }))+g(p_2,
\epsilon_2({\beta })),\end{equation}
We present them with the color factor $\delta_{cd}$ for the final-state
gluons removed.  That is, we write the amplitude
${\cal A}(i,j;f)$ in Eq.~(\ref{H}) as
\begin{equation}
{\cal A}(i,j;f) = \delta_{cd}{\bar A}(i,j;\alpha, \beta).
\end{equation}
Note also that to get ${\cal A}$ normalized as in Eq.~(\ref{mdp})
we contracted the incoming gluons' color indices
with $\delta^{ab}/(N^2_c-1) = \delta_{ab}/8$.
Our notation for the polarizations ($\epsilon_{1'}(i)$,
$\epsilon_{1}(\alpha)$,
etc.)\ indicates that we are using a different set of
basis states for each of the four gluons.  A gluon's
polarization vector must be transverse with respect to its
momentum: $\epsilon_p\cdot p=0$.

For the incoming gluons, we define linear polarization vectors as
\begin{eqnarray}
\epsilon_{1'}(1)&=&\epsilon_{2'}(1)=(0,0;1,0)\nonumber\\
\epsilon_{1'}(2)&=&\epsilon_{2'}(2)=(0,0;0,1),\end{eqnarray}
and for the outgoing gluons, we define polarization vectors
in ($I$) and out ($O$)
of the $x$-$y$ plane as
\begin{eqnarray}
\epsilon_1(I)&=&\epsilon_2(I)=(0,0;\sin\phi ,-\cos
\phi ),\nonumber\\
\epsilon_1(O)&=&\sqrt {\frac {ax_a}{2bx_b}}\left(1,\frac {
-bx_b}{ax_a};{\bf 0}\right),\nonumber\\
\epsilon_2(O)&=&\sqrt {\frac {bx_a}{2ax_b}}\left(1,\frac {
-ax_b}{bx_a};{\bf 0}\right).\end{eqnarray}

Then straightforward Feynman graph calculations give:
\begin{eqnarray}
{\bar A}(1,1,I,I)&=&{\bar A}(2,2,O, O)
=ig^2\frac 34\left[\left(\frac 4{\kappa}-1\right)\sin^2\phi +\cos^
2\phi\right],\nonumber\\
{\bar A}(1,1,O,O)&=&{\bar A}(2,2,I,I)
=ig^2\frac 34\left[\left(\frac 4{\kappa}-1\right)\cos^2\phi +\sin^
2\phi\right],\nonumber\\
{\bar A}(1,2,I,O)&=&{\bar A}(2,1,O,I)
=-ig^2\frac 34\left[\frac 1a\cos^2\phi +\frac 1b\sin^2\phi\right
],\nonumber\\
{\bar A}(1,2,O,I)&=&{\bar A}(2,1,I,O)
=ig^2\frac 34\left[\frac 1b\cos^2\phi +\frac 1a\sin^2\phi\right
],\nonumber\\
{\bar A}(1,1,I,O)&=&{\bar A}(1,1,O,I)
=-{\bar A}(2,2,I,O)=-{\bar A}(2,2,O
,I),\nonumber\\
&=&ig^2\frac 34\left[\frac 1a-\frac 1b\right]\sin\phi\cos\phi ,\nonumber\\
{\bar A}(1,2,I,I)&=&{\bar A}(2,1,I,I)
=-{\bar A}(1,2,O,O)=-{\bar A}(2,1,O
,O)\nonumber\\
&=&ig^2\frac 32\left[1-\frac 2{\kappa}\right]\sin\phi\cos\phi .\nonumber\\
\label{ampls}\end{eqnarray}

\end{document}